# Evaluation of Key Impression of Resilient Supply Chain Based on Artificial Intelligence of Things (AIoT)

**Alireza Aliahmadi[1], Hamed Nozari[2,*]** 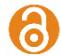**, Javid Ghahremani-Nahr[3], Agnieszka Szmelter-Jarosz[4]**

[1] Department of Management and Industrial Engineering, Iran University of Science and Technology, Tehran, Iran; aaliahmadi@iust.ac.ir.
[2] Department of Industrial Engineering, Iran University of Science and Technology, Tehran, Iran; ham.nozari.eng@iauctb.ac.ir.
[3] Faculty Member of Academic Center for Education, Culture and Research (ACECR), Tehran, Iran; javid.ghahremanii@yahoo.com.
[4] Department of Logistics, Faculty of Economics, University of Gdańsk, Poland; agnieszk.szmelterjarosz@gmail.com.

## Abstract

In recent years, the high complexity of the business environment, dynamism and environmental change, uncertainty and concepts such as globalization and increasing competition of organizations in the national and international arena have caused many changes in the equations governing the supply chain. In this case, supply chain organizations must always be prepared for a variety of challenges and dynamic environmental changes. One of the effective solutions to face these challenges is to create a resilient supply chain. Resilient supply chain is able to overcome uncertainties and disruptions in the business environment. The competitive advantage of this supply chain does not depend only on low costs, high quality, reduced latency and high level of service. Rather, it has the ability of the chain to avoid catastrophes and overcome critical situations, and this is the resilience of the supply chain. AI and IoT technologies and their combination, called AIoT, have played a key role in improving supply chain performance in recent years and can therefore increase supply chain resilience. For this reason, in this study, an attempt was made to better understand the impact of these technologies on equity by examining the dimensions and components of the Artificial Intelligence of Things (AIoT)-based supply chain. Finally, using nonlinear fuzzy decision making method, the most important components of the impact on the resilient smart supply chain are determined. Understanding this assessment can help empower the smart supply chain.

## 1 | Introduction

In recent years, due to the importance of supply chain in creating a competitive advantage, competition between organizations has become competition between their supply chains. Therefore, strong supply chain management is a key factor in the success of organizations. A variety of developments as a result of unexpected events are an integral part of the supply chains of today's organizations that operate in conditions of uncertainty. A resilient approach to supply chain management has been developed to deal with these sudden breakdowns in the supply chain and rapid recovery and return to the original state before the event. Resilience refers to the ability of the supply chain to cause unexpected perturbations, and the ability of the system to return to its baseline state or to a more favorable state after a perturbation [1].

There are two types of risks in each supply chain: operational risks and disruption risks. Operational risks relate to inherent uncertainties such as demand, supply, waiting times for delivery, prices, availability of raw materials, quantity and quality of returned products; therefore, the need to discover the random nature of supply chains is a very important concern [2]. On the other hand, given that the likelihood of disruption risks such as sanctions, currency fluctuations and inflation is high for a productive and knowledge-based organization. Therefore, in such circumstances, the need to design a resilient supply chain pattern becomes more apparent. Therefore, the supply chain must be prepared to face any event and, while providing an efficient and effective response, be able to return to the original state; this is the meaning of supply chain resilience. Resilience is at the heart of supply chain management thinking. Supply chain organizations need to implement and guide the use of resources by implementing an appropriate supply chain resilience assessment system so that they can more properly control and manage their path to achieve the desired goals while respecting the environmental consequences [3].

In the real world, identifying components alone is not enough, as they are directly or indirectly related to each other and have a degree of interactive relationship. Therefore, there is a need for an approach that can identify the internal relationships between components and key dimensions and analyze the effect of one variable on other variables and consider the intensity of the effect of one variable on other variables based on the real thought of individuals. Also, supply chain management must move towards different, innovative and technological approaches in order to be more capable in dealing with risk disorders. One of these approaches is resilient supply chain [4]. The presence of digital technologies in today's world has transformed all aspects of abandoned business. Supply chain as one of the most important parts of business has been affected by these technologies such as Internet of Things (IoT) technology and Artificial Intelligence (AI) and its sub-sectors such as Machine Learning (ML). The IoT is one of the most important sources of big data production, and if the power of AI analyzer is added to it, Artificial Intelligence of Things (AIoT) technology is obtained which can add more predictive and analytical power to supply chains. This capability can create solutions to deal with environmental turbulence and crises [5]. Using this approach can increase the organization's ability to reduce costs and challenges that occur suddenly. Considering the importance of using these technologies and the growing growth of these technologies and their amazing effects in smartening supply chain processes, in this study, the dimensions and components of a resilient intelligent supply chain based on AIoT technology were examined. Also, a nonlinear non-linear fuzzy decision method was used to evaluate the key components of resilient and intelligent amin chains. The results show the importance of each of the key dimensions and effective components. Using this framework and evaluation to powerfully implement a lean supply chain can be effective.

The rest of the research is organized in this way. The Section 2 provides a literature review. In Section 3, the dimensions and components of the resilient intelligent supply chain are introduced. In Section 4, the research method is specified. In the Section 5, the results are presented and finally, the discussion and conclusion will be expressed.

## 2 | Literature Review

Many definitions of the concept of resilience have been expressed by various researchers. But in general, in most studies, the concept of resilience is defined as "the degree of resilience of systems to crises". This concept is used in a number of disciplines including economics, politics, engineering and planning. The purpose of creating resilience in the supply chain is to prevent the chain from moving towards unfavorable conditions and to restore the supply chain after the disruption occurs in the shortest time and at the lowest cost [6]. In addition, supply chain resilience should not be seen as merely the ability to manage risk, but the ability to respond to risk in a better and more cost-effective way than other competitors, and ultimately to gain a competitive advantage. Supply chain resilience can be described as the ability of a supply chain to return to its original state or move to a new or even more desirable state after being disrupted. In other words, the ability of the supply chain to prepare for unforeseen events, respond to disruptions and recover them by maintaining the continuity of operations at the desired level, continuity and monitoring of its structure and performance, supply chain resilience is defined [7].

Vugrin et al. [8] examined the concept of supply chain resilience. According to these researchers, the concept of resilience increases investment in various aspects of the supply chain, including adaptability, flexibility and recyclability. Despite divergent and ambiguous definitions of supply chain resilience, there are commonalities in these definitions:

− *Flexibility, which is the ability of infrastructure systems to minimize disruptions in the event of a disaster and optimally improve it before disruption occurs.*
− *Resilience enhancement strategies that are defined in three time periods before the occurrence of the disorder, at the time of the occurrence of the disorder and after the onset of shock and disorder.*

Therefore, managers should not focus on tragic events (which may occur); rather, it is necessary to manage and develop the supply chain in such a way that the chain can respond well to progressive disruptions and return to normal. The ability to respond appropriately to natural or man-made disasters is a strategic need for an organization to survive in a competitive environment; especially when the organization is part of a network of entities [9]. Resilience is a category that needs to be designed. Flexible supply chains are able to adapt effectively to turbulence to maintain the same level of efficiency. Today, transformational technologies are tools that help businesses on the path to resilience. AIoT depend on the skills of the data, tools and software that keep the supply chain running. Supply chain flexibility is increased through AIoT applications. Sensor devices that track inventory movement among retailers, distributors and customers provide clear guidance on issues such as shipping delays. Analysis of this data provides high predictive power in the face of sudden situations. Tools designed to monitor workforce skills also use ML to predict whether staff shortages or skills may affect operations [10]. Many researchers have studied the resilience characteristics of the supply chain. Soni [11] discuss the characteristics of resilient supply chains in terms of flexibility, compatibility, collaboration, visibility and stability. These features make competitive profit possible and also reduce the level of disruption in supply chains. Prior to the introduction of the IoT in the supply chain, data and information were shared with only one actor, which reduced this transparency, unlike the smart supply chain, which allows all required actors to work simultaneously [12]. Visibility was also a problem due to the lack of real-time data and information, which led to errors, inaccuracies and information distortions throughout the supply chain. Complexity and uncertainty create many problems in the supply chain. For example, mistakes in the delivery and production of new orders that lead to a waste of time and high costs for business owners. The goal of implementing evolving technologies such as the combination of AIoT and IoT has changed the business model in traditional supply chains, leading to the redesign of past business models [13]. The characteristics of the AIoT-based resilient supply chain are shown in *Fig. 1*.

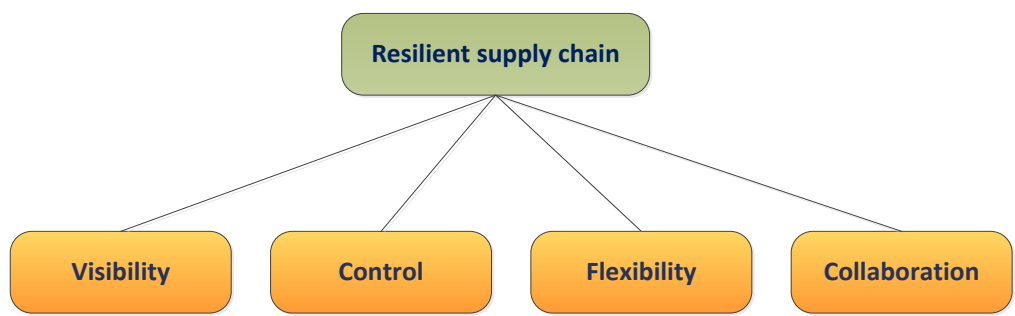

Fig. 1. Characteristics of AIoT-based resilient supply chain [14].

The inclusion of these technologies in processes reduces costs, complexity and inaccuracy along the supply chain. Understanding supply chain processes requires information sharing and data transparency across the chain. The presence of AIoT technologies as well as technologies such as blockchain increases trust in the supply chain. Therefore, because of the effective and timely response, the risks are minimized.

Collaboration requires access to information between relevant partners. In a resilient supply chain, ensuring the reduction of ambiguity and event preparation is critical, as resilient supply chain processes are extensive [15]. Resilience ensures that all kinds of developments and crises can be dealt with without a negative impact on the supply chain. Resilience involves the speed and agility of various functions in the supply chain [16].

The IoT can be introduced to add real-time aspects and improve supply chain flexibility performance. Data mining, cloud computing, AI-based analysis, and ML are some of the methods that go beyond controlling traditional supply chain technologies by controlling and tracking supply chain performance. Real-time identification and traceability are the fastest ways in which data and information are transmitted across sections and delivered to supply chain actors to help make decisions in emergencies or sudden disruptions [17]. Therefore, it can be concluded that the use of smart technologies can affect the supply chain from different dimensions to ensure a resilient supply chain. IoT and blockchain technology enhance supply chain visibility through enhanced transparency across the supply chain, enabling the accuracy and reliability of the data sent. Therefore, with the presence of AI technology along with IoT technology, support for the decision-making process and the trust of consumers and partners will increase [18]. Intelligent technologies can also enhance supply chain resilience by providing a stable chain and disaster tolerance using programmable processes such as intelligent scheduling methods. Information from sensors and other IoT tools provide intelligent control over the supply chain using real-time monitoring and control. This ensures security throughout the chain. In addition, improved product and service management increases the ability to track resources in real time. Accuracy in inventory control increases supply chain agility by speeding up flow processes and information analysis. Thus, AIoT offers promising potential for handling supply chain resilience, error handling speed, and robust big-data analysis [19].

## 3 | Dimensions and Components of Resilient Smart Supply Chain

Every year, natural disasters around the world cause a lot of damage, the number of which has increased in recent years and has significant human, structural and economic effects. The study of resilience is therefore essential to understanding how business ecosystems respond to major harms, process and economic challenges and changes in international business. In fact, in the face of such a crisis, supply chain managers must be able to have a complete picture of their supply chain to assess risks and prepare scenarios for predictable issues related to human, material, technical or financial. The IoT, along with platforms that support functional analytics, AI and ML (AIoT-based intelligent supply chains) can help, create transparency from manufacturers to retailers, and raise awareness. Increase efficient decision making. By providing a detailed view of the products. All this makes the supply network more responsive and flexible in the face of changing market conditions [20].

The new RFID and GPS sensors can track products "from floor to store", giving the company an overview of what is happening in the supply chain at the commodity level. Command centers centralize information in one place, provide better visibility, and help companies have more precise control over quality control, timely delivery, and product forecasting. Some command centers are even equipped with software that makes quick settings on the fly easy.

For this reason, and due to the high importance of this issue, in this study, an attempt has been made to review the literature as well as the opinions of experts on the dimensions and key impacts of an smart resilient supply chain based on AIoT technology. *Table 1* shows the most important dimensions and components affecting the resilience of an intelligent supply chain based on these evolving technologies.

Table 1. Dimensions and impression of smart resilient supply chain based on AIoT technology.

| Dimension | Code | Impression | Code | Reference |
|---|---|---|---|---|
| Triggers | C1 | Risk reduction | C11 | [18, 21, 22] |
| | | Uncertainty reduction | C12 | [23, 24, 25] |
| Vulnerabilities | C2 | Environmental turbulence | C21 | [26, 27] |
| | | Connections | C22 | [28, 22] |
| Capabilities | C3 | flexibility | C31 | [29, 30] |
| | | Transparency | C32 | [31, 32] |
| Empowerment | C4 | Responsiveness | C41 | [33, 18] |
| | | Agility | C42 | [29, 22] |
| | | Risk management | C43 | [34, 35] |
| | | power of prediction | C44 | [36, 26, 18] |
| | | Performance improvements | C45 | [37, 18] |

**Triggers.** There are factors that cause the need for resilience in the process. The presence of big data and the extraction of this data throughout the supply chain process and their analysis can prevent many risks.

**Vulnerabilities.** Factors are a fundamental impediment that predisposes the company to disruption and its ability to withstand threats as well as to survive in the event of accidental events (both inside and outside the system).

**Capabilities.** Features that enable a company to anticipate and overcome disruptions. These features are essential for delivering the right performance and success.

**Empowerment.** Factors that strengthen the supply chain resilience.

## 4 | Research Method

This research uses a quantitative method to evaluate the effective dimensions and impressions in an smart resilient supply chain based on AIoT technology. A fuzzy nonlinear hierarchical analysis technique was used to prioritize the effective components in the resilient supply chain. For this analysis, experts active in the field of supply chain in FMCG industries (due to the authors' access) were used. In this study, Cronbach's alpha method was used to assess the reliability of the relevant questionnaires using SPSS software. The Cronbach's alpha value of all questionnaires was 0.821 in total, which is considered desirable. *Fig. 2* shows the research methodology.

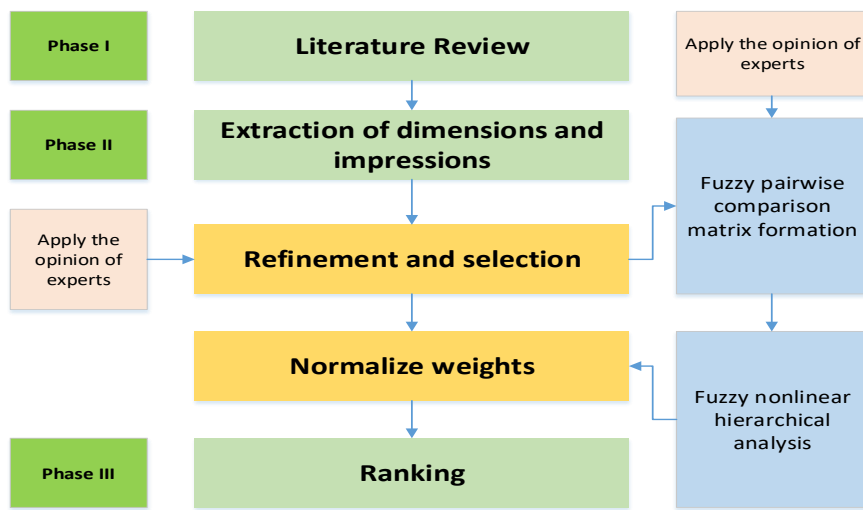

Fig. 3. Research framework.

In this research, a fuzzy nonlinear hierarchical analysis method, which is representative of Mikhailov method [38], has been used to prioritize the effects of intelligent resilient supply chain. The steps for using this method are as follows:

**Drawing the hierarchical structure.** Shown in *Table 1*.

**Formation of fuzzy pairwise comparison matrix.** Fuzzy judgment agreement matrices are formed based on the opinions of experts. For this reason, fuzzy numbers have been used to express the preferences of experts (FMCG industry supply chain). The linguistic variables and their fuzzy scale are presented in *Table 2*.

Table 2. Linguistic variables for pairwise comparisons.

| Linguistic Variable | Triangular Fuzzy Scale |
|---|---|
| very Low | (1,2,3) |
| Low | (2,3,4) |
| medium | (3,4,5) |
| High | (4,5,6) |
| Very high | (5,6,7) |

As shown in *Table 2*, fuzzy triangular numbers are used for linguistic scales.

**Modeling and problem solving.** The definite weight vector $w = (w_1, w_2, \ldots, w_n)$ is extracted in such a way that the priority rate is approximately within the range of the initial fuzzy judgments. That is, the weights are determined so that *Eq. (1)* is established.

$$l_{ij} \leq \frac{w_i}{w_j} \leq u_{ij}. \qquad (1)$$

Each definite weight vector (w) holds with a degree in the above fuzzy inequalities obtained by the linear membership function of *Eq. (2)*.

$$\mu_{ij}\left(\frac{w_i}{w_j}\right) = \begin{cases} \dfrac{(w_i/w_j) - l_{ij}}{m_{ij} - l_{ij}} & \dfrac{w_i}{w_j} \leq m_{ij} \\ \dfrac{u_{ij} - (w_i/w_j)}{u_{ij} - m_{ij}} & \dfrac{w_i}{w_j} \leq m_{ij} \end{cases}. \qquad (2)$$

Considering the specific form of membership functions, the fuzzy prioritization problem is presented to a nonlinear optimization problem in the form of *Model (3)*.

$$\begin{aligned}
&\max \lambda \\
&\text{Subject to:} \\
&(m_{ij} - l_{ij})\lambda w_j - w_i + l_{ij} w_j \leq 0, \\
&(u_{ij} - m_{ij})\lambda w_j + w_i - u_{ij} w_j \leq 0, \\
&i = 1, 2, \ldots, n-1 \quad j = 2, 3, \ldots, n \quad j > i \\
&\sum_{k=1}^{n} w_k = 1. \quad w_k > 0 \quad k = 1, 2, \ldots, n
\end{aligned} \qquad (3)$$

Positive optimal values for the index, indicate that all weight ratios are completely true in the initial judgment, but if the index is negative, it can be seen that the fuzzy judgments are strongly inconsistent and the weight ratios are almost true in these judgments.

## 5 | Research Finding

The steps for evaluating and ranking the dimensions and impressions of an AIoT-based intelligent resilient supply chain are divided into two main parts:

– *Determining the matrix of pairwise comparisons based on the integration of experts' opinions.*
– *Using mathematical modeling in order to rank and gain the weight of impressions in the research model.*

In order to evaluate and prioritize in this study, fuzzy questionnaires using language variables were sent to experts active in the FMCG industry. These pairwise comparison tables are shown in *Tables 3* to *7*. These tables have been used for non-linear prioritization calculations.

Table 3. Matrix of pairwise comparisons for dimensions.

|     | C1  |     |     | C2  |     |     | C3  |     |     | C4  |     |     |
| --- | --- | --- | --- | --- | --- | --- | --- | --- | --- | --- | --- | --- |
| C1  | -   | -   | -   | -   | -   | -   | -   | -   | -   | -   | -   | -   |
| C2  | 3.2 | 4.2 | 5.2 | -   | -   | -   | -   | -   | -   | -   | -   | -   |
| C3  | 2.4 | 2.8 | 4.7 | 2.5 | 3.1 | 4.5 | -   | -   | -   | -   | -   | -   |
| C4  | 3.2 | 3.5 | 5.5 | 3.2 | 3.5 | 4.7 | 2.5 | 2.75| 3.65| -   | -   | -   |

Table 4. Paired comparison matrix for triggers.

|     | C11 |     |     | C12 |     |     |
| --- | --- | --- | --- | --- | --- | --- |
| C11 | -   | -   | -   | -   | -   | -   |
| C12 | 2.7 | 3.47| 4.25| -   | -   | -   |

Table 5. Paired comparison matrix for vulnerabilities.

|     | C21 |     |     | C22 |     |     |
| --- | --- | --- | --- | --- | --- | --- |
| C21 | -   | -   | -   | -   | -   | -   |
| C22 | 1.25| 2.75| 3.25| -   | -   | -   |

Table 6. Paired comparison matrix for capabilities.

|     | C31 |     |     | C32 |     |     |
| --- | --- | --- | --- | --- | --- | --- |
| C31 | -   | -   | -   | -   | -   | -   |
| C32 | 2.11| 3.8 | 4.5 | -   | -   | -   |

Table 7. Matrix of pairwise comparisons for empowerment.

|     | C41 |     |     | C42 |     |     | C43 |     |     | C44 |     |     | C45 |     |     |
| --- | --- | --- | --- | --- | --- | --- | --- | --- | --- | --- | --- | --- | --- | --- | --- |
| C41 | -   | -   | -   | -   | -   | -   | -   | -   | -   | -   | -   | -   | -   | -   | -   |
| C42 | 2.5 | 3.2 | 5.1 | -   | -   | -   | -   | -   | -   | -   | -   | -   | -   | -   | -   |
| C43 | 1.25| 3.1 | 4.9 | 3.1 | 3.25| 5.11| -   | -   | -   | -   | -   | -   | -   | -   | -   |
| C44 | 1.75| 2.5 | 5.24| 1.1 | 2.75| 5.1 | 1.5 | 3.75| 4.12| -   | -   | -   | -   | -   | -   |
| C45 | 2.11| 3.21| 4.14| 2.1 | 2.75| 3.27| 1.75| 2.2 | 4.1 | 3.1 | 3.5 | 4.8 | -   | -   | -   |

By placing the data from *Tables 3* to *7* in the nonlinear *Model (3)* and modeling using the LINGO software, the weight and rank of each of the criteria and impressions can be obtained. The computational results are shown in *Tables 8* to *12*.

Table 8. Weight and ranking of dimensions.

| Dimension       | Code | Weight   | Rank | $\lambda$ |
| --------------- | ---- | -------- | ---- | --------- |
| Triggers        | C1   | 0.112142 | 4    |           |
| Vulnerabilities | C2   | 0.180125 | 3    | 0.41025   |
| Capabilities    | C3   | 0.253274 | 2    |           |
| Empowerment     | C4   | 0.455241 | 1    |           |

**Table 9. Weight and ranking of impressions related to triggers.**

| Impressions | Code | Weight | Rank | $\lambda$ |
|---|---|---|---|---|
| Risk reduction | C11 | 0.458452 | 2 | 0.25412 |
| Uncertainty reduction | C12 | 0.554126 | 1 | |

**Table 10. Weight and ranking of impressions related to vulnerabilities.**

| Impressions | Code | Weight | Rank | $\lambda$ |
|---|---|---|---|---|
| Environmental turbulence | C21 | 0.518472 | 1 | 0.32145 |
| Connections | C22 | 0.488216 | 2 | |

**Table 11. Weight and ranking of impressions related to capabilities.**

| Impressions | Code | Weight | Rank | $\lambda$ |
|---|---|---|---|---|
| flexibility | C31 | 0.429987 | 2 | 0.27415 |
| Transparency | C32 | 0.572145 | 1 | |

**Table 12. Weight and ranking of impressions related to empowerment.**

| Impressions | Code | Weight | Rank | $\lambda$ |
|---|---|---|---|---|
| Responsiveness | C41 | 0.174152 | 3 | |
| Agility | C42 | 0.165214 | 4 | |
| Risk management | C43 | 0.243214 | 2 | 0.41721 |
| power of prediction | C44 | 0.292142 | 1 | |
| Performance improvements | C45 | 0.122415 | 5 | |

As shown in *Tables 8* to *12*, a positive value for the compatibility index indicates the acceptable compatibility of the matrices. By normalizing the weights, we can get their total weight and overall rank. The normalized computational results are shown in *Table 13*.

**Table 13. Normal weight and rank of dimensions and impressions of AIoT-based resilient supply chain.**

| Dimension | Weight | Impressions | Weight | Normalized Weight | Rank |
|---|---|---|---|---|---|
| Triggers | 0.112142 | Risk reduction | 0.458452 | 0.051412 | 11 |
| | | Uncertainty reduction | 0.554126 | 0.062141 | 9 |
| Vulnerabilities | 0.180125 | Environmental turbulence | 0.518472 | 0.093390 | 5 |
| | | Connections | 0.488216 | 0.087940 | 6 |
| Capabilities | 0.253274 | flexibility | 0.429987 | 0.108905 | 4 |
| | | Transparency | 0.572145 | 0.144909 | 1 |
| Empowerment | 0.455241 | Responsiveness | 0.174152 | 0.079281 | 7 |
| | | Agility | 0.165214 | 0.075212 | 8 |
| | | Risk management | 0.243214 | 0.110721 | 3 |
| | | power of prediction | 0.292142 | 0.132995 | 2 |
| | | Performance improvements | 0.122415 | 0.055728 | 10 |

*Fig. 3* shows the rank and position of the effects of the intelligent resilient supply chain based on IoT technologies and AI.

As can be seen from *Table 13*, the capability dimension is the most important dimension of the resilient supply chain based on intelligent technologies. Transparency is also the biggest impact of this AIoT-based supply chain, which can be reasonable given the presence of IoT technology and pure data.

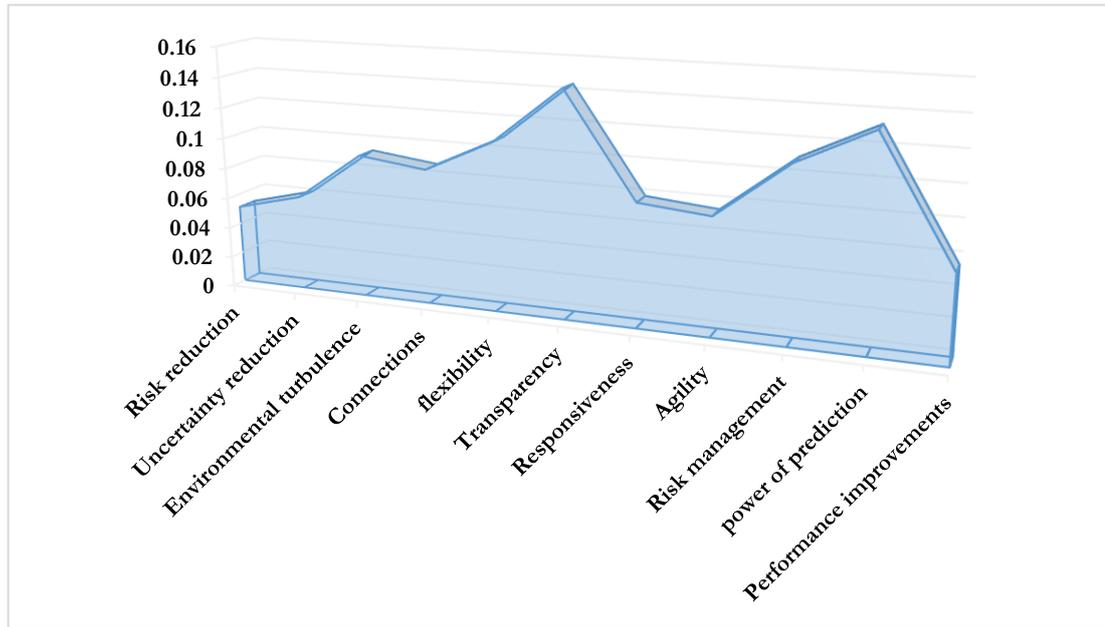

**Fig. 3. Rank and position of AIoT-based resilient supply chain effects.**

## 6 | Conclusion

A resilient supply chain may not be the least expensive supply chain, but a resilient supply chain can overcome uncertainties and disruptions in the business environment. The competitive advantage of the supply chain does not depend only on low costs, high quality, reduced latency and high level of service. Rather, it has the ability of the chain to avoid catastrophes and overcome critical situations, and this is the resilience of the supply chain. Resilience is the ability of the supply chain to overcome unpredictable events. The purpose of supply chain resilience is to prevent the chain from moving toward adverse conditions. Accountability and recovery to the same level or better is a common feature of all resilience-related perspectives, including environmental, social, psychological, economic, organizational, and crisis management. Maintaining the same level of control (similar to the normal level) on structure and performance in the event of a disturbance is an essential feature of environmental resilience. The IoT is a powerful technology that has had a tremendous impact on business processes in recent years. In addition to being the largest source of big data production, the technology also helps to create pure and secure data. The presence of this technology along with AI technology creates the powerful AIoT technology, which also has high analytical power and helps to understand and react quickly to many sudden situations.

For these reasons, and considering the importance of the presence of this technology and its high effects, in this study, an attempt was made to study the dimensions and impressions of the presence of these technologies for supply chain resilience. For this purpose, using literature review, the most important dimensions and impressions of technology presence in supply chain resilience were extracted. Then, using a nonlinear decision making method, these dimensions and effects were prioritized. The results show that transparency and strong predictability are among the most important effects of the AIoT-based resilient supply chain, which makes sense given the presence of technologies such as the IoT.

## Conflicts of Interest

All co-authors have seen and agree with the contents of the manuscript and there is no financial interest to report. We certify that the submission is original work and is not under review at any other publication.